\documentclass[journal]{IEEEtran}
\usepackage[left=2cm,right=2cm,top=2cm,bottom=3cm]{geometry}
\usepackage{amsmath}
\usepackage{rotating}
\usepackage{color}
\usepackage{longtable}
\usepackage{array,makecell}
\usepackage{verbatim} 

\usepackage[switch]{lineno} 
\usepackage{lipsum}

\newcommand{\bi}{\begin{itemize}}
\newcommand{\ei}{\end{itemize}}
\newcommand{\bd}{\begin{displaymath}}
\newcommand{\ed}{\end{displaymath}}
\newcommand{\be}{\begin{equation}}
\newcommand{\ee}{\end{equation}}
\newcommand{\bea}{\begin{eqnarray}}
\newcommand{\eea}{\end{eqnarray}}
\newcommand{\ba}{\begin{array}}
\newcommand{\ea}{\end{array}}
\newcommand{\bc}{\begin{center}}
\newcommand{\ec}{\end{center}}

\begin{document}
{\huge \textbf{IEEE copy right notice}}

©2020 IEEE.  Personal use of this material is permitted.  Permission from IEEE must be obtained for all other uses, in any current or future media, including reprinting/republishing this material for advertising or promotional purposes, creating new collective works, for resale or redistribution to servers or lists, or reuse of any copyrighted component of this work in other works.

Accepted to be published in: IEEE Transactions on Neural Systems and Rehabilitation Engineering

\newpage
\title{The influence of posture, applied force and perturbation direction on hip joint viscoelasticity}

\author{Hsien-Yung Huang, Arash Arami, Ildar Farkhatdinov, Domenico Formica and Etienne Burdet
\thanks{Email:\{h.huang14,e.burdet\}@imperial.ac.uk. All authors are or were with the Department of Bioengineering, Imperial College of Science, Technology and Medicine, UK. Arami is with the Department of Mechanical and Mechatronics Engineering, University of Waterloo, Canada. Farkhatdinov is with the School of Electronics Engineering and Computer Science, Queen Mary University of London, UK. Formica is with the Department of Engineering, University Campus Bio-Medico di Roma, Italy. We thank Jonathan Eden for editing the manuscript. This work was funded in part by the EU-FP7 grants ICT-601003 BALANCE and ICT-611626 SYMBITRON.}}

\maketitle

\begin{abstract}
Limb viscoelasticity is a critical factor used to regulate the interaction with the environment. It plays a key role in modelling human sensorimotor control, and can be used to assess the condition of healthy and neurologically affected individuals. This paper reports the estimation of hip joint viscoelasticity during voluntary force control using a novel device that applies a leg displacement without constraining the hip joint. The influence of hip angle, applied limb force and perturbation direction on the stiffness and viscosity values was studied in ten subjects. No difference was detected in the hip joint stiffness between the dominant and non-dominant legs, but a small dependency was observed on the perturbation direction. Both hip stiffness and viscosity increased monotonically with the applied force magnitude, with posture to being observed to have a slight influence. These results are in line with previous measurements carried out on upper limbs, and can be used as a baseline for lower limb movement simulation and further neuromechanical investigations.

\end{abstract}
\section{Introduction}
\label{ch:Introduction}
Muscles are characterised by their viscoelasticity, where stiffness and viscosity increase with activation. By co-activating the muscles acting on limbs, the human nervous system can control its stiffness and viscosity in magnitude, shape and orientation \cite{Burdet2013}. Critically, this enables humans to regulate their interaction with the environment \cite{Hogan1985} e.g. during object manipulation, or for running optimally on different grounds. 

In order to understand how humans control the limb viscoelasticity, a large body of experiments have estimated stiffness and viscosity in the upper limb, in particular at the wrist and arm \cite{Burdet2013}. Stiffness and viscosity can be measured indirectly by applying a mechanical disturbance on the limb and regressing the resulting changes of position and force. Measurements carried out using this method showed that stiffness generally increases linearly with the applied force: in one deafferented muscle, in a single joint (thus including reflexes), and in the arm \cite{Burdet2013}.

Much less is known on the viscoelasticity in the lower limbs, in part due to the difficulty to carry out suitable experiments involving heavy leg mass. For instance, existing robotic interfaces to estimate viscoelasticity in the lower limb either require a sitting or lying position \cite{Amankwah2004, Sinclair2006, Perell1996, Akman1999}, or are not sufficiently rigid to apply fast perturbations without causing non-negligible oscillations e.g. \cite{Lunenburger2005, Koopman2016, Meuleman2016, Farkhatdinov2019}. In addition, all of these interfaces are affixed to the body and thus determine the joints around which the limb can move, while anatomical joints generally vary with the posture (e.g. the knee joint rotates and translates during locomotion). An alternative method consists of applying perturbations directly on the foot, which can be used to estimate ankle viscoelasticity \cite{Lee2011,Rouse2014}.

In view of the limitations of previous devices to investigate the lower limb viscoelasticity, we have developed a dedicated robotic interface \cite{Huang2019a, Huang2019c}. This rigid interface can be used to investigate the hip, knee or ankle neuromechanics in a natural upright posture. It uses an endpoint-based approach to apply dynamic environments on the leg, thus does not need to impose joint movement.

Due to the difficulty to apply a mechanical disturbance on the leg for estimating viscoelasticity, experiments reported in the literature have been mainly restricted to a single joint, i.e. at the ankle \cite{Mirbagheri2000} and knee joints \cite{Pfeifer2012, Ludvig2017}. In \cite{Koopman2016} the LOPES exoskeleton has been used to estimate viscoelasticity at the whole leg (including the hip joint), using a multi-joint random torque as perturbation and an indirect measurement of the resulting displacement from its series elastic actuators. Random torque perturbations enable experiments to identify both the stiffness and viscosity simultaneously \cite{Perreault2001, Perreault2002}, but may lead to identification problems as the velocity dependent component is much smaller than the position dependent component \cite{Gomi1996,Lee2015,Lee2016}. Therefore, we preferred using a single position displacement to focus on accurately determining the joint stiffness \cite{Burdet2000}, and estimated viscosity in a second step using the whole perturbation, including the ramp up before the constant displacement phase and the ramp down after it. This allowed us to examine the effect of individual factors such as posture or force level separately.

\section{Methods}
\label{ch:Methods}
\subsection{Measurement system}
\label{ch:Measurement system}
The Neuromechanics Evaluation Device (NED) is a powerful cable-driven robotic interface to yield computer-controlled dynamic testing on one leg of subjects supported in a seated or upright posture (\cite{Huang2019a}, Fig.\ref{fig:NEDSketch}a). NED's open stand support allows for conducting biomechanics identification experiments on various subjects including subjects with impaired motor function. Used in different configurations, this cable-based system can control the motion of the whole leg, foreleg, or foot in order to estimate the hip, knee or ankle neuromechanics. The pulley system can be adjusted to keep the cable orientation approximately normal to the limb’s movement in different orientations for subjects of various size \cite{Huang2019a}.

\subsection{Experiment}
\label{ch:Experiment}
The experimental protocol was approved by the Imperial College Research Ethics Committee. Safety measures with NED include software limits on the velocity, acceleration and jerk, an optical system to check perturbation limits, and emergency buttons for the subject and experimenters \cite{Huang2019a}. Ten subjects (of age 21-27, with 6 females) without any known lower-limb injury or medical condition were recruited, they were informed on the device and experiment and signed a consent form prior to participation. Subjects' weight and leg length (from the anterior superior iliac spine to the lateral malleolus) were then measured to estimate leg inertia. These subjects' parameters are reported in Table \ref{tab:TableOfSubjects}. 

Bipolar electromyography (EMG) electrodes placed on the rectus femoris, biceps femoris and tibialis anterior muscles were used to check when subjects are relaxed. EMG signals were recorded at 2048 Hz, filtered using a [5,500]Hz bandpass Butterworth filter, followed by a notch filter at 50 Hz (to attenuate the power frequency), then rectified. A locking knee brace was used to keep the knee joint fixed during the perturbations, and thus ensure that the leg is straight during the whole procedure.
\begin{table}[!h]
\centering
\caption{Biographical information of the subjects}
\label{tab:TableOfSubjects}
\begin{tabular}{|p{0.03\columnwidth}|p{0.16\columnwidth}|p{0.15\columnwidth}|p{0.2\columnwidth}|p{0.04\columnwidth}|p{0.04\columnwidth}|}
\hline
{\bf no} & {\bf weight [kg]} & {\bf height [m]} & {\bf leg length [m]} & {\bf age} & {\bf sex}\\ \hline
1 & 67 & 1.70 & 0.89 & 25 & M\\ \hline
2 & 47 & 1.55 & 0.82 & 24 & F\\ \hline
3 & 100 & 1.79 & 0.85 & 27 & M\\ \hline
4 & 47 & 1.55 & 0.82 & 26 & F\\ \hline
5 & 61 & 1.72 & 0.93 & 23 & F\\ \hline
6 & 54 & 1.68 & 0.88 & 27 & F\\ \hline
7 & 54 & 1.72 & 0.94 & 21 & F\\ \hline
8 & 69 & 1.71 & 0.85 & 25 & M\\ \hline
9 & 85 & 1.79 & 0.87 & 23 & M\\ \hline
10 & 50 & 1.50 & 0.81 & 24 & F\\ \hline
\end{tabular}
\end{table}

\begin{figure}[]
\centering
\includegraphics[width=\linewidth]{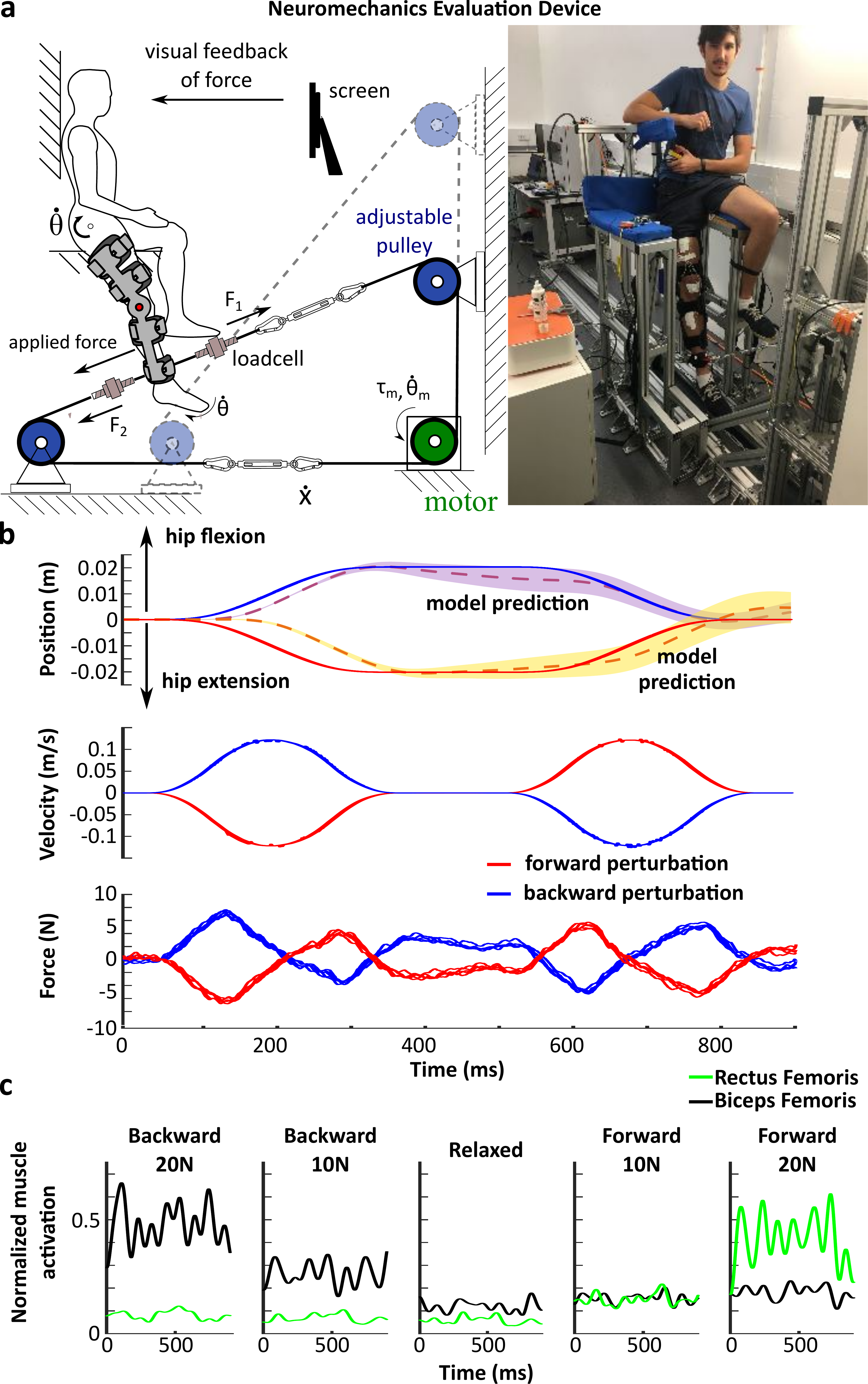}
\caption[Sketch of NED and perturbation profile example]{\small Sketch of Neuromechanics Evaluation Device (NED) and perturbation profile used to estimate the hip viscoelasticity. Panel (a) depicts an the experimental setup. The subject's leg was moved by the motor via a cable closed loop. The interaction force was recorded by the loadcells in both the front and back of the ankle. The pulleys can be displaced to yield a force perpendicular to the subject's leg. $\dot{\theta}_m$ and $\tau_m$ are the speed and torque at the motor, $\dot{\theta}$ the hip joint angular velocity, $\dot{X}$ the cable linear motion, $F_1$ and $F_2$ the force recorded at the loadcell in both front and the back. Visual feedback of the applied force enabled the subjects to control a desired force level while a perturbation was provided by the interface. Panel (b) shows the measured position, velocity and interaction force $\delta$($F_1 - F_2$) of a representative trial. The displacement predicted with the linear model using the data of one subject and condition is shown as dashed line with 95\% confidence area.} Panel (c) shows the EMG envelope computed using a Butterworth low-pass filter with 10 Hz cut off frequency and normalized with the maximum value (in the 20N condition). 
\label{fig:NEDSketch}
\end{figure}

Each participant was asked to relax while supporting their body weight using the handle. A harness was used to connect the ankle of the leg under test to the cable system (Fig.\ref{fig:NEDSketch}a). The subject could familiarise with the device by experiencing several perturbations, after which the system workspace safety limits were set.

A position perturbation was used to estimate hip joint impedance. The perturbation shown in Fig.\ref{fig:NEDSketch}b was used. It consists of a 150ms long plateau with 20mm amplitude (corresponding e.g. to an angle of 1.15$^\circ$ for a 90cm long leg) with smooth ramps up and down. This perturbation profile was determined by trial and error to ensure a force measurement profile with negligible oscillations \cite{Huang2019a}. All data but the EMG was measured at 1000 Hz.

For both legs, measurement was carried at different initial postures with the hip angle (relative to vertical) at \{15$^\circ$, 25$^\circ$, 35$^\circ$, 45$^\circ$, 55$^\circ$\}. At every posture, subjects were first asked to relax (which was checked using EMG) while a perturbation (with profile as in Fig.\ref{fig:NEDSketch}b) was applied by the system randomly in the forward or backward direction, with five trials in each direction. The time of a perturbation was also random so that the subject could not prepare for a perturbation.

After experiencing the perturbation while relaxed (0N) condition, each subject was asked to pull or push the leg to exert a force of \{-20, -10, 10, 20\}N (with positive value for backward kick) as was controlled by the subject using real-time feedback of the applied force displayed on a computer screen placed in front of them. The force level was taken relative to the relaxed condition of each subject, so that the effect of gravity was compensated by the interface. To prevent a subject from volitionally reacting to a perturbation, visual feedback was not updated during the perturbation. 
\begin{figure*}[]
\centering
\includegraphics[width= 1.9\columnwidth]{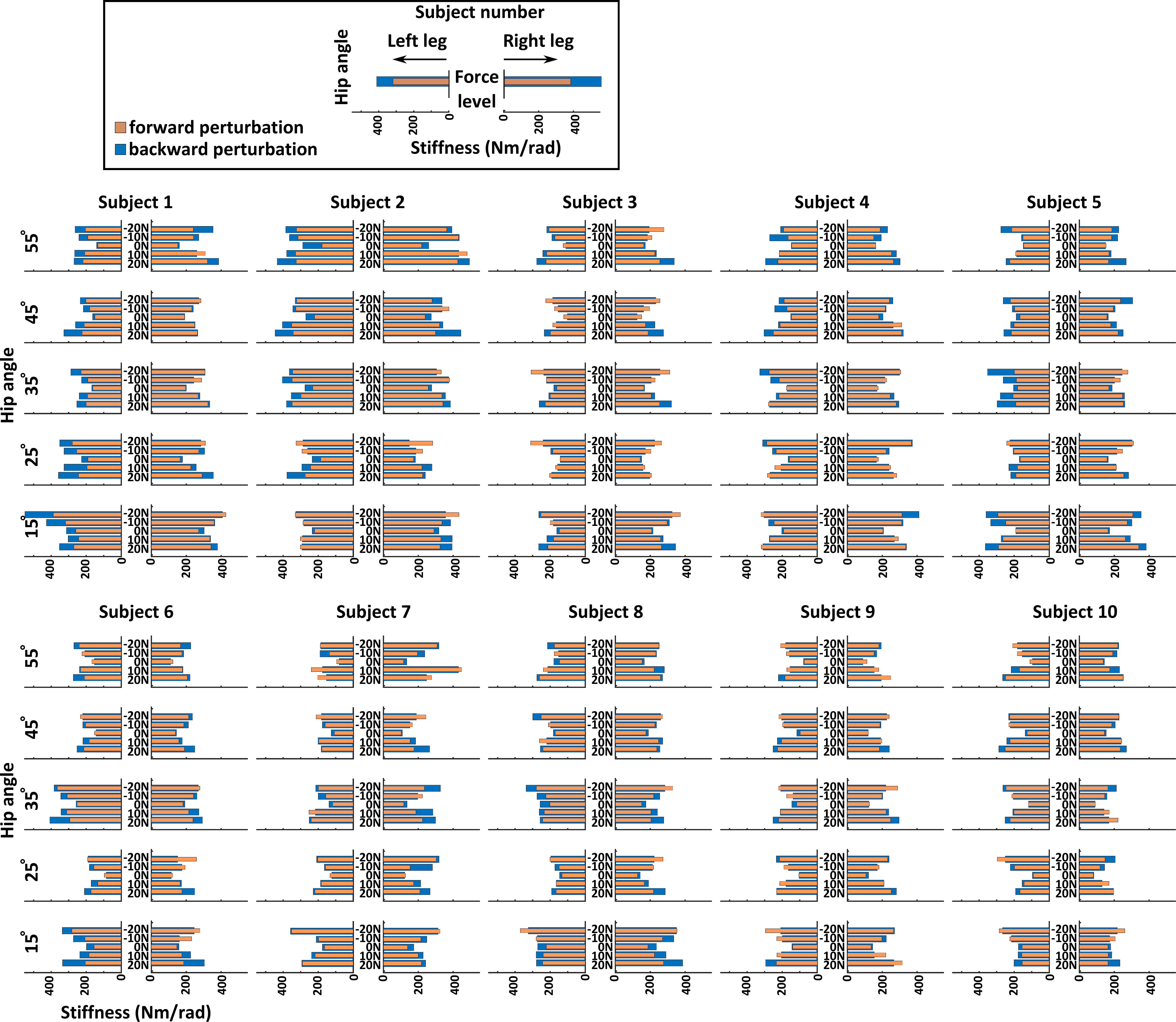}
 \caption[Estimated stiffness of all participants]{\small Hip stiffness results for all subjects and conditions.}
\label{fig:AllData}
\end{figure*}

The subjects carried out two such measurement cycles (5 minutes each), with a ten minute rest during which they were detached from NED. For the two legs of the ten subjects, there were thus ten trials at each of the five postures and five force levels, using two perturbation directions (see Fig.\ref{fig:AllData}). The total experiment time was 100 minutes excluding the breaks. 

\subsection{Data analysis}
\label{ch:Data analysis}
Linearization of the hip joint dynamics (valid for small angles $\delta \theta$) yields
\begin{equation}
\label{eq:model_all_CH3}
\tau_m + \tau_g + \tau = I \delta \ddot{\theta} + B \delta \dot{\theta} + K \delta \theta\, ,
\end{equation}
where $\tau_m$ is a torque produced by muscle tension to counteract the gravitational torque $\tau_g$ and $\tau$ corresponding to the external forces. $I \delta \ddot{\theta}$ is the inertia component and $B \delta \dot{\theta} + K \delta \theta$ the hip viscoelasticity component corresponding to a displacement angle $\delta \theta$, where $B$ is the viscosity and $K$ stiffness. 

Two loadcells are used at the extremities of the ankle fixture to record the interaction forces between the robot and the subject. As the change of gravitational torque $\delta \tau_g$ between the extremities was found to have less than 1\% effect on the overall joint torque with the 20mm perturbation amplitude, it is considered as negligible. Furthermore, $\tau_m$ can be considered as constant, as the visual feedback was not updated during perturbation thus there is no reaction to the perturbation (as shown in Fig.\ref{fig:NEDSketch}b). The dynamics measured by the loadcells is thus:
\begin{equation}
\label{eq:jointimpedance_CH3}
L \, \delta(F_1 \! - \! F_2) \equiv \delta\tau = I\delta\ddot{\theta}+B\delta\dot{\theta}+K\delta\theta \, ,
\end{equation}
where $L$ is the leg length, $F_1$ and $F_2$ are the forces measured at the front and rear loadcells, respectively, and $\delta \tau$ is the torque response to $\delta \theta$.

Similar to the method described in \cite{Burdet2000}, a constant displacement (as shown in Fig.\ref{fig:NEDSketch}b) was used to identify stiffness $K$ using: 
\begin{equation}
\label{eq:stiffness_CH3}
\delta\tau \equiv \, K \, \delta\theta \, .
\end{equation}
For each participant, leg, posture, force level, and perturbation direction condition, the perturbation displacement $\delta \theta$ and resulting change of torque $\delta \tau$ in the last 100ms of the perturbation plateau of all 10 trials formed 1x1000 vectors, which were used to estimate $K$ as the least-square solution of Eq.\ref{eq:stiffness_CH3}.

Viscosity was determined in a second step as the least-square solution of the transfer function (using Matlab {\it tfest} command with search method set 'auto' for best fit):
\begin{equation}
\label{eq:jointimpedance_CH3_Part2}
 \frac{\Delta \Theta (s)}{\Delta T (s)} = \frac{1}{Is^2 + Bs + K},
\end{equation}
where $\Delta \Theta$ and $\Delta T$ are the Laplace transforms of $\delta \theta$ and $\delta \tau$ respectively. In this equation, inertia was computed from the biomechanical model of \cite{Winter1995} and stiffness was estimated from Eq.\ref{eq:stiffness_CH3}. The weight of the leg was estimated as 16.1\% of total weight, and the radius of gyration of the whole leg at the distal end is 0.56$L$, thus
\begin{equation}
\label{eq:Inertia}
I = 0.161 M (L \, 0.56)^2,
\end{equation}
with the mass $M$ and length $L$ parameters from Table \ref{tab:TableOfSubjects}. For each subject, the angle ($\delta \theta$) and torque data ($\delta \tau$) over the whole perturbation period (900ms) for the five trials corresponding to a specific (posture, force level and perturbation direction) condition are used together to identify viscosity using Eq.\ref{eq:jointimpedance_CH3_Part2}. Instead of concatenating the data of all five trials, it is "grouped" and defined as multi-experiment data (using Matlab {\it merge}) to avoid potential prediction error due to the transition period between two concatenated time series data.

The quality of the identification was tested by predicting the displacement from the force data and the identified values for $K,I,B$. As can be seen in Fig.\ref{fig:NEDSketch}b, the prediction generally follows that actual displacement. The delay of the predicted position probably stems from delays in the force measurement due e.g. to cable compliance. The mean and standard deviation of the calculated coefficient of determination ($R^2$) in all subjects are listed in Table.\ref{tab:TableOfRSquare}. 

\begin{table}[!h]
\centering
\caption{Reliability of model prediction}
\label{tab:TableOfRSquare}
\begin{tabular}{|p{0.25\columnwidth}|p{0.25\columnwidth}|p{0.25\columnwidth}|}
\hline
subject number & mean R-square & SD of R-square\\ \hline
 1 & 0.58 & 0.14 \\ \hline
 2 & 0.55 & 0.19 \\ \hline
 3 & 0.59 & 0.18 \\ \hline
 4 & 0.57 & 0.11 \\ \hline
 5 & 0.54 & 0.24 \\ \hline
 6 & 0.49 & 0.16 \\ \hline
 7 & 0.67 & 0.15 \\ \hline
 8 & 0.53 & 0.14 \\ \hline
 9 & 0.39 & 0.19 \\ \hline
 10 & 0.47 & 0.39\\ \hline

\end{tabular}
\end{table}

\section{Results}
\label{ch:Results}
Fig.\ref{fig:AllData} summarizes the stiffness estimation results of all ten subjects. These results were obtained with the two perturbation directions, for their two legs, at the selected five postures and the five force levels. Hip joint overall stiffness changes with the perturbation direction, applied limb force level and hip angle (as was tested by separate Friedman's tests with p$<$0.05). No difference was detected between stiffness values in the dominant and non-dominant legs (as was tested using both Friedman's test and paired t-test). In the following, we will investigate how stiffness and viscosity depend on the perturbation direction, force level and hip posture.

\begin{figure}[]
\centering
\includegraphics[width= 0.9\columnwidth]{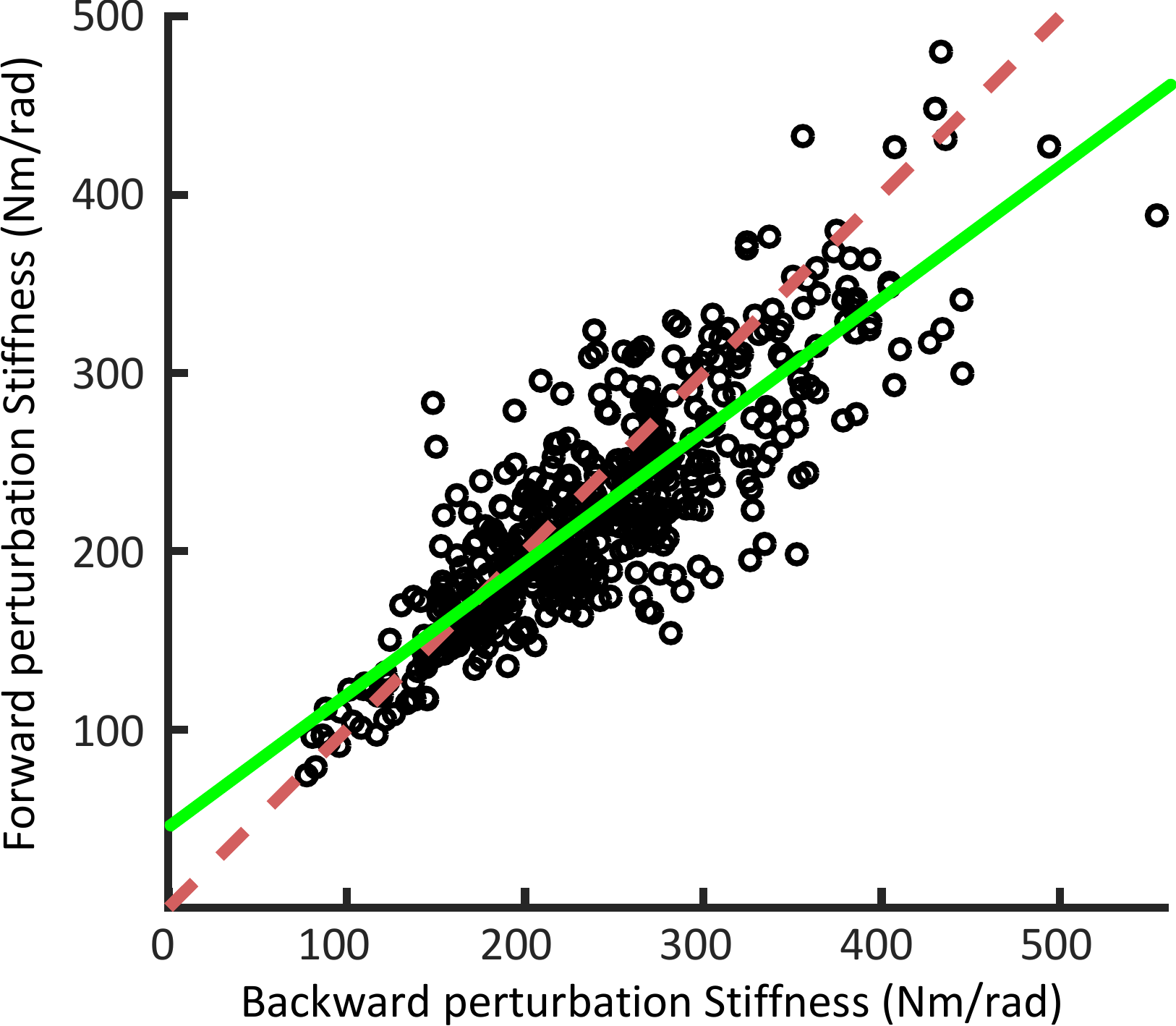}
\caption[Hip stiffness measurement depends on the perturbation direction.]{\small Hip stiffness measurement depends on the perturbation direction. Each dot represents the stiffness at a specific subject leg, posture and force level, with stiffness measured with backward perturbation in the abscissa and with forward perturbation in the ordinate. The linear regression (green solid line) below the (dashed red) diagonal indicates larger values with perturbations in backward as in forward directions.}
\label{fig:TwoDirectionPerturbation}
\end{figure}

{\it Perturbation direction dependency.} Fig.\ref{fig:TwoDirectionPerturbation} shows how the stiffness values of all subjects, at all postures and force levels, depend on the perturbation direction. We see that a larger portion of the stiffness values is below the identity line, suggesting that the backward perturbation results in larger stiffness values than the forward perturbation. This was confirmed by a paired t-test indicating that the difference between the estimation was different with the two different directions (p$<$0.05). The linear regression result (green solid line, with R$^2$=0.72) described in Table~\ref{tab:MixedEffectModel} exhibits a difference of 26\% between the estimation in the two directions. On the other hand, the estimated viscosity values showed no clear perturbation direction dependency, with regression close to identity line but R$^2\!<$0.1 for the best linear regression model.

\begin{figure}[]
\centering
\includegraphics[width=\columnwidth]{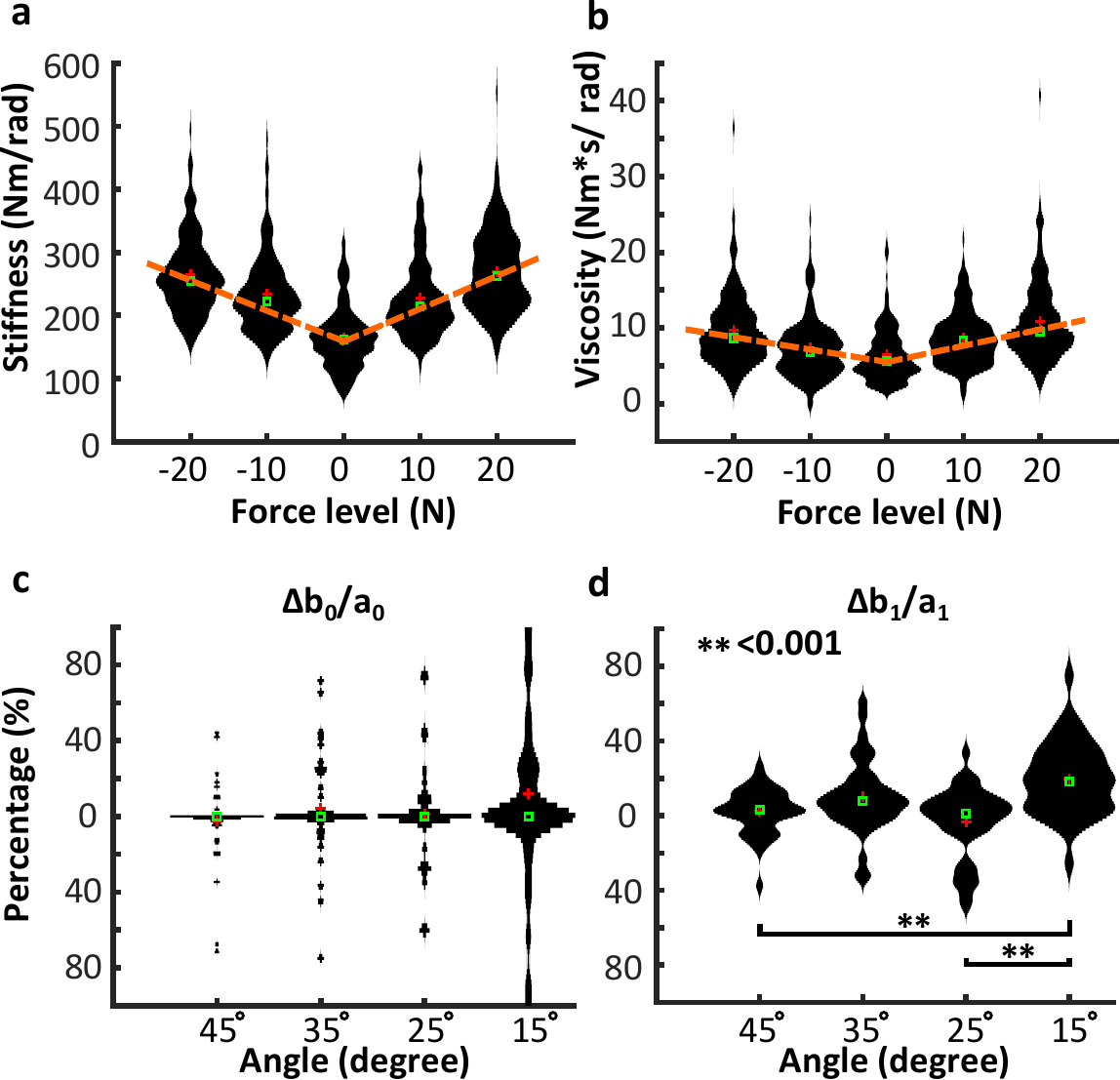}
\caption[Force level dependency and changes due to hip angle]{\small Violin plots showing the probability density of force-level dependency and how it changes due to hip angle. The dashed lines indicate the least square fitted force dependency. Panels (a) and (b) show how hip stiffness and viscosity changes with the applied force. Panel (c) and panel (d) shows the influence of hip angle upon force-level dependency. The influence is presented as random effects ($b_{0i}$ and $b_{1i}$) and specifically in the percentage of fixed effects ($a_0$ and $a_1$). Additionally, it is presented as changes with respect to hip angle 55$^\circ$ in order to examine changes from a specific hip angle. Within each violin plots, a cross indicates the median value of the respective violin plot and a square the mean value. Random effects are found to change the identified force-level dependency ($a_0$) inconsistently and does not reach statistically significant at any hip angle. On the other hand, relaxed stiffness ($a_1$) is found to change with hip angle and confirmed to be statistical significant by two tailed Wilcoxon rank sum test and corrected by Bonferroni correction.}
\label{fig:ForceDependencyAndRandomEffects}
\end{figure}

{\it Force-level dependency.}
To investigate the relationship between measured viscoelasticity, applied limb force level and hip angle, we performed three steps of mixed effect modelling to examine the stiffness change due to the selected parameter. Firstly, stiffness was assumed to vary linearly with applied limb force while posture may influence this linear relation, modeled as:
\bea
\label{eq:stiffness_force}
\mathbf{K} \!\!\!\!&=&\!\!\! \mathbf{X}\, \boldsymbol{\beta} + \mathbf{Z} \, \boldsymbol{\mu} + \boldsymbol{\varepsilon} \\
\begin{bmatrix} 
\!\!K_{11} \!\!\\ \vdots \\ \!\! K_{1m} \!\!\\ \!\!K_{21}\!\! \\ \vdots \\ \!\!K_{2m} \!\!\\ \vdots \\ \!\!K_{n1}\!\! \\ \vdots \\ \!\!K_{nm} \!\!
\end{bmatrix} \!\!\!\!\!\!\!&\equiv&\!\!\!\!\!\!\!
\begin{bmatrix} 
\!\!F_{11} \!\!\!&\!\!\! 1 \!\\ 
\vdots \!\!\!&\!\!\! \vdots \\ 
\!\!F_{1m} \!\!\!&\!\!\! 1 \!\\ 
\!\!F_{21} \!\!\!&\!\!\! 1 \!\\ 
\vdots \!\!\!&\!\!\! \vdots \\ 
\!\!F_{2m} \!\!\!&\!\!\! 1 \!\\ 
\vdots \!\!\!&\!\!\! \vdots\\ 
\!\!F_{n1} \!\!\!&\!\!\! 1 \!\\
\vdots \!\!\!&\!\!\! \vdots\\ 
\!\!F_{nm} \!\!\!&\!\!\! 1 \!
\end{bmatrix}
\!\!\!\begin{bmatrix}
a_0\!\\ a_1 \!\end{bmatrix} \!\! + \!\!
\begin{bmatrix} 
\!\!F_{11} \, 1 & \!\!\!0\,\,\,\,\,0 & \hdots & 0\,\,\,\,\,0\! \\ 
\vdots & \!\!\!\vdots & & \\ 
\!\!F_{1m} \, 1 & \!\!\!0\,\,\,\,\,0 & & \\ 
\!\!0\,\,\,\,\,0 & \!\!\!F_{21}\,1 & & \vdots \\ 
 & \!\!\!\vdots & & \\ 
 & \!\!\!F_{2m}\,1 & & \\ 
\vdots & \!\!\!0\,\,\,\,\,0 & \ddots & 0\,\,\,\,\,0\! \\
 & \!\!\!\vdots & & F_{n1} \, 1 \!\\
 & \!\!\!\vdots & & \vdots \\
\!\!0\,\,\,\,\,0 & \!\!\!0\,\,\,\,\,0 & & F_{nm} \, 1 \!
\end{bmatrix}\!\!\!
\begin{bmatrix}
b_{01}\!\!\\ b_{11} \!\!\\ b_{02}\!\!\\ b_{12}\!\! \\ \vdots \\ b_{0n}\!\! \\ b_{1n}\!\!
\end{bmatrix}
\!\!\! + \!\boldsymbol{\varepsilon}
\nonumber
\eea
where $\mathbf{K}$ are the stiffness values of one subject's leg measured at $n=5$ different hip angles and $m=6$ force and perturbation factors. $m=6$ corresponds to $3$ (either positive, or negative) force levels, and $2$ perturbation directions. \{$\mathbf{X}, \boldsymbol{\beta}$\} are to capture the fixed effect and $\{\mathbf{Z}, \boldsymbol{\mu}\}$ to test the influence of hip angle upon the identified force-stiffness relation. $\boldsymbol{\varepsilon}$ is the error.
By estimating mixed effect models for each subject's leg, it was found that stiffness increases monotonically with applied force amplitude in all subjects (presented in Fig.\ref{fig:ForceDependencyAndRandomEffects}a). The estimated force-level dependency weight ($a_0$) has a mean value of 5.15Nm/rad per applied Newton force and a standard deviation of 0.98Nm/rad. This finding indicates a positive relationship between applied limb forces and hip joint stiffness, which is further confirmed by F-tests (p$<$0.05 for all subjects' legs).

Furthermore, Friedman tests showed that the hip angle would change both fixed-effect parameters, namely the relaxed stiffness ($a_1$) and force-level dependency ($a_0$) (with p=0.0006 and p$<$0.0001, respectively). To further emphasize stiffness change due to hip angle, random effects are presented as the relative percentage of fixed effects ($b_{0i}/a_0$ and $b_{1i}/a_1$). Furthermore, the acquired percentages were further subtracted by random effect percentages estimated at 55$^\circ$ hip angle in order to present stiffness change with respect to 55$^\circ$ hip angle. As shown in Fig.\ref{fig:ForceDependencyAndRandomEffects}d, relaxed stiffness ($a_1$) changes with posture and reached statistically significance at 15$^\circ$ degree hip angle (tested with two tailed Wilcoxon rank sum test with Bonferroni correction). On the other hand, Fig.\ref{fig:ForceDependencyAndRandomEffects}c shows that the force-level dependency ($a_0$) changed inconsistently due to posture and does not reach statistical significance at any specific hip angle.

The same investigation was carried out on the estimated viscosity. All subjects had an increased viscosity with applied force (with a mean slope of 0.19Nm\,s/rad, presented in Fig.\ref{fig:ForceDependencyAndRandomEffects}b). However, only 42\% cases passed the F-test, indicating that the viscosity change due to the applied limb force may be insignificant. Additionally, the identified mixed effect models showed low prediction accuracy and a limited data variance explained by the model (with mean $\overline{R^2}$=0.35 over the subjects lower than the stiffness model prediction with R$^2$=0.79) despite the inclusion of random effects. It is, therefore, unclear whether hip joint viscosity exhibits similar force-level dependency or whether the identified trend was merely due to noise.

{\it Posture dependency.}
To better catch the larger stiffness at 25$^\circ$ and 15$^\circ$, a second investigation used a model assuming that stiffness changes quadratically with hip angle. We tested how the applied limb force may influence this quadratic relation by using the following model:
\bea
\label{eq:stiffness_angle}
\mathbf{K} \!\!\!\!&=&\!\!\!\!\!
\begin{bmatrix} 
\theta_{11}^2 \!\!\!&\!\!\! \theta_{11} \!\!\!&\!\!\! 1 \\
\vdots \!\!\!&\!\!\! \vdots \!\!\!&\!\!\! \vdots \\
\theta_{nm}^2 \!\!\!&\!\!\! \theta_{nm} \!\!\!&\!\!\! 1 \\
\end{bmatrix} \!\!\!
\begin{bmatrix} a_2 \\ a_3 \\ a_4 \end{bmatrix} \!+ \mathbf{Z}\, \boldsymbol{\mu} + \boldsymbol{\varepsilon} \\
\boldsymbol{\mu} \!\!\!\!&\equiv&\!\!\!\! 
\begin{bmatrix} 
b_{21}\, \,
b_{31}\,\,
b_{41}\,\,
b_{22}\, \,
b_{32}\, \,
b_{42}\,
\hdots\,
b_{2n}\, \,
b_{3n}\, \,
b_{4n}
\end{bmatrix}^T \nonumber
\eea
where K are the stiffness values measured at $n=5$ different force and $m=10$ angle and perturbation factors. $m=10$ corresponds to $5$ hip angles and $2$ perturbation directions. The random-effect design matrix $\mathbf{Z}$ is built based on elements of the fixed-effect design matrix $\mathbf{X}$, and can be found in the Appendix~\ref{ch:app}. $\boldsymbol{\mu}$ is the random-effect vector indicating the influence of the ith applied force on the stiffness and hip angle relation, where $\{b_{2i}\}$ show the influence on the quadratic angle term and $\{b_{3i}\}$ on the linear angle term, and $\{b_{4i}\}$ demonstrate the constant effect.

The identified fixed effect parameters indicated that most legs exhibit an inverse relationship between measured stiffness and hip angle, as presented in Table~\ref{tab:MixedEffectModel} in combination with F-test results. In other words, it was found within our experiment range \{15-55$^\circ$\} that hip joint stiffness would increase with the decrease of hip angle. However, the identified posture dependency was less influential compared to the previously identified force-level dependency, as the model without random effects showed a low estimation accuracy (mean over the subjects $\overline{R^2}$= 0.16) and required random effects that consider applied limb forces (mean over the subjects $\overline{R^2}$= 0.65). The importance of force-level dependency was consolidated by theoretical likelihood tests (where 95\% cases passed with p$<$0.05), and suggested that applied limb force is a stronger influencing factor in comparison with hip angle.

The same process was repeated on estimated viscosity with Eq.\ref{eq:stiffness_angle}. The identified models showed poor prediction accuracy and explained limited variance of data (mean over the subjects $\overline{R^2}$=0.28) with 65\% of the models failed the F-tests (indicating no posture dependency).
\begin{figure}[]
\centering
\includegraphics[width= 0.95\columnwidth]{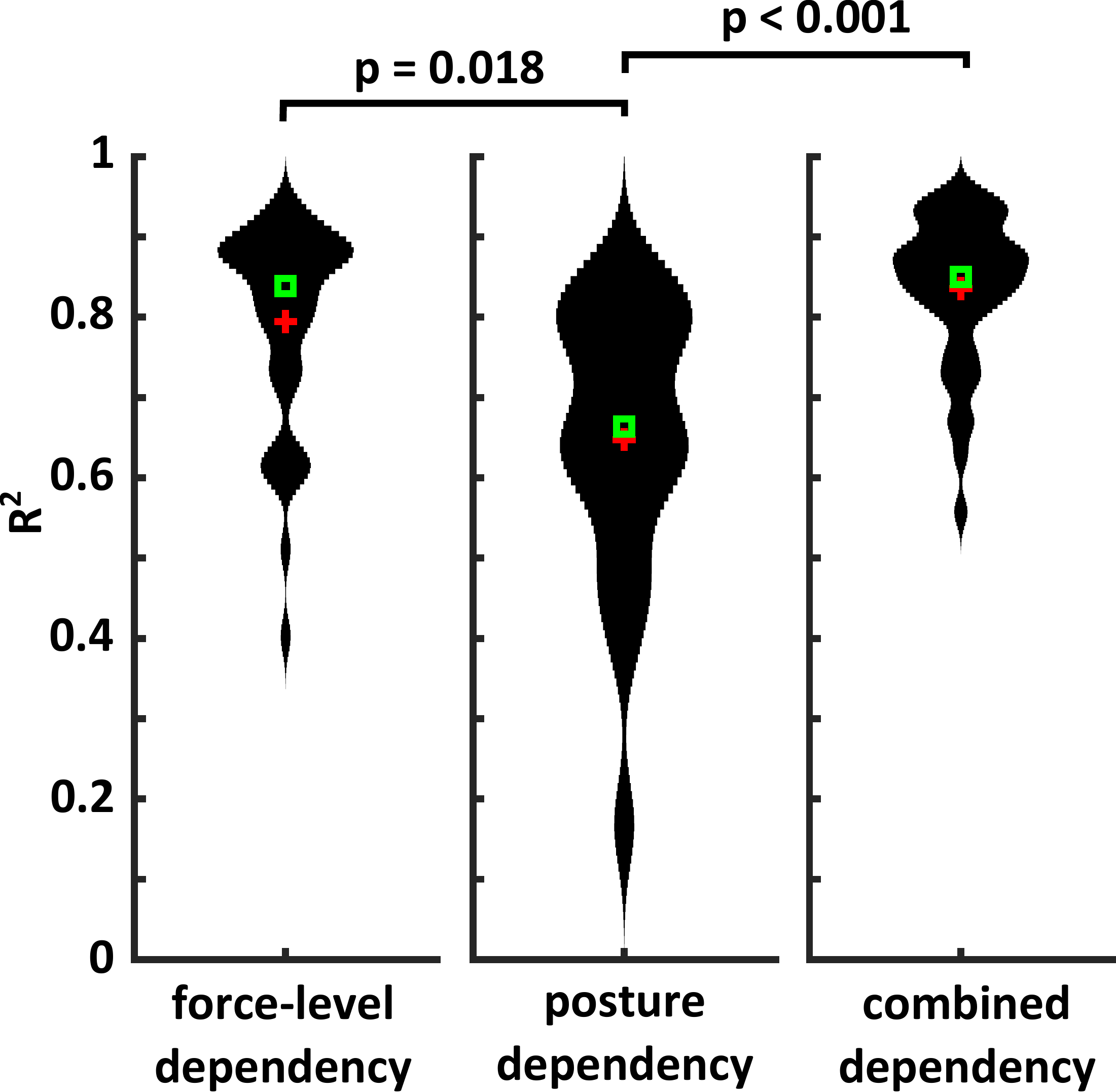}
\caption[Model prediction accuracy comparison]{\small Model prediction accuracy comparison. Prediction accuracy is presented as R$^2$ and compared between all three models. It is shown that both models that considers force-level dependency performed a better prediction (tested with two tailed Wilcoxon rank sum test). On the other hand, the combined model improves estimation accuracy, however, did not reach a statistical significant (with p = 0.3579).}
\label{fig:ModelComparison}
\end{figure}

{\it Force and posture dependency.}
Based on the aforementioned test results, we further hypothesised that stiffness changes according to both applied limb force and hip angle, with each factor possibly affecting the other one:
\bea
\label{eq:stiffness_angle+force}
\mathbf{K} \!\!\!\!&=&\!\!\!\!\!
\begin{bmatrix} 
F_{11} \!\!&\!\! \theta_{11}^2 \!\!&\!\! \theta_{11} \!\!&\!\! 1 \\
\vdots \!\!&\!\! \vdots \!\!&\!\! \vdots \!\!&\!\! \vdots \\
F_{nm} \!\!&\!\! \theta_{nm}^2 \!\!&\!\! \theta_{nm} \!\!&\!\! 1 \nonumber \\
\end{bmatrix} \!
\begin{bmatrix} a'_0 \\ a'_2 \\ a'_3 \\ a_5 \end{bmatrix} \!+ \mathbf{Z}_1 \boldsymbol{\mu}_1 + \mathbf{Z}_2 \boldsymbol{\mu}_2 + \boldsymbol{\varepsilon} \\
\boldsymbol{\mu}_1 \!\!\!\!&\equiv&\!\!\!\! 
\begin{bmatrix} 
b'_{01}\, \,
b'_{11}\, \,
b'_{02}\, \,
b'_{12}\, \,
\hdots\,
b'_{0n}\, \,
b'_{1n}
\end{bmatrix}^T  \\ 
\boldsymbol{\mu}_2 \!\!\!\!&\equiv&\!\!\!\! 
\begin{bmatrix} 
b'_{21}\, \,
b'_{31}\, \,
b'_{41}\, \,
b'_{22}\, \,
b'_{32}\, \,
b'_{42}\, \,
b'_{23}\, \,
b'_{33}\, \,
b'_{43}
\end{bmatrix}^T \nonumber
\eea
where K are the stiffness values measured at $n=5$ different angle and $m=6$ force and perturbation factors. $m=6$ corresponds to $3$ (either positive, or negative) force levels and $2$ perturbation directions. The random-effect design matrices $\mathbf{Z}_1$ and $\mathbf{Z}_2$ are built based on elements of the fixed-effect design matrix $\mathbf{X}$ and can also be found in Appendix~\ref{ch:app}. $\boldsymbol{\mu}_1$ and $\boldsymbol{\mu}_2$ are the random-effect vectors with $\boldsymbol{\mu}_1$ indicating the influence of the {\it i}-th hip angle on the stiffness and force relation, and $\boldsymbol{\mu}_2$ indicating the influence of applied force on the angle relation.

Interestingly, the newly identified fixed effects exhibited values similar to previous findings. Stiffness was again found to increase with applied limb force, with slopes (mean $a_0'$=4.98) close to previous values (mean $a_0$=5.15). By calculating the differences between both values, 83\% cases showed differences less than 10\% (calculated by $(a'_0 - a_0)/a_0$). Meanwhile, most subjects were again found to exhibit a negative relation between stiffness and hip angle, and are presented in Table~\ref{tab:MixedEffectModel} along with F-test results. These findings imply that the identified force-level and posture dependencies coexist.

The estimated generalised linear models, which refers to models without random effects, were shown to predict hip joint stiffness of all subjects' legs with acceptable variance being explained (mean over the subjects $\overline{R^2}$=0.68, with standard deviation of 0.16). The model can be further improved by including random effects (mean over the subjects $\overline{R^2}$=0.84, with standard deviation 0.09, 92.5\% cases passed F-tests). This finding demonstrates the importance of correlation among parameters (e.g. hip angle changing force-level dependency). On the other hand, random effects ($b_{1i}'$ and $b_{4i}'$) which affect the constant value ($a_5$) are shown to decrease since both posture and force-level dependencies are considered in this model.

The model prediction accuracy of all three models is presented in Figure~\ref{fig:ModelComparison}.

\begin{table}[]
\centering
\caption{Statistics of linear regression and mixed effect models}
\label{tab:MixedEffectModel}
\begin{tabular}{|p{0.35\columnwidth}|p{0.18\columnwidth}|p{0.3\columnwidth}|}
\hline
{} & expected value & standard deviation \\ \hline \hline
\multicolumn{3}{|c|}{{\bf Stiffness: perturbation direction dependency}} \\ \hline
\multicolumn{3}{|c|}{ $Y = 0.74X+45.29, \quad R^2=0.717$ } \\ \hline
{intercept} & 45.29 & 5.26  \\ \hline 
{slope} & 0.74 & 0.02  \\ \hline \hline
\multicolumn{3}{|c|}{{\bf Stiffness: force level dependency} (Eq.\ref{eq:stiffness_force}, $\overline{R^2}=0.79$)} \\ \hline
{a$_0$} [m/rad] & 5.15 & 0.98  \\ \hline 
{a$_1$} [Nm/rad] & 169.39 & 39.61  \\ \hline 
{b$_{0i}$/a$_0$} & 0 & 19.93\%  \\ \hline 
{b$_{1i}$/a$_1$} & 0 & 14.24\%  \\ \hline 
\multicolumn{3}{|c|}{\thead{Identified dependencies: 100\% cases found force-level dependency}} \\ \hline
\hline
\multicolumn{3}{|c|}{{\bf Stiffness: posture dependency} (Eq.\ref{eq:stiffness_angle}, $\overline{R^2}=0.84$)} \\ \hline
{a$_2$} [Nm/rad$^3$] & 137.12 & 240.90  \\ \hline 
{a$_3$} [Nm/rad$^2$] & -212.96 & 303.65  \\ \hline 
{a$_4$} [Nm/rad]& 302.75 & 104.06  \\ \hline 
{b$_{2i}$/a$_2$} & 0 & 21.41\%  \\ \hline 
{b$_{3i}$/a$_3$} & 0 & 6.1\%  \\ \hline 
{b$_{4i}$/a$_4$} & 0 & 14.19\%  \\ \hline 
\multicolumn{3}{|c|}{\thead{Identified dependencies: \\
20\% cases failed F-tests, showing no posture dependency \\
5\% cases showed positive posture dependency\\
75\% cases showed negative posture dependency}} \\ \hline
\hline
\multicolumn{3}{|c|}{{\bf Stiffness: posture and force-level dependency} (Eq.\ref{eq:stiffness_angle+force}, $\overline{R^2}=0.84$)} \\ \hline
a$_0'$ [m/rad] & 4.98 & 1.34  \\ \hline 
a$_2'$ [Nm/rad$^3$] & 117.58 & 230.51  \\ \hline 
a$_3'$ [Nm/rad$^2$] & -186.59 & 292.69  \\ \hline 
{a$_5$} [Nm/rad] & 234.00 & 102.24  \\ \hline 
b$_{0i}'$/a$_0'$ & 0 & 15.97\%  \\ \hline 
b$_{1i}'$/a$_5$ & 0 & 13.11\%  \\ \hline 
b$_{2i}'$/a$_2'$ & 0 & 31.46\%  \\ \hline 
b$_{3i}'$/a$_3'$ & 0 & 5.02\%  \\ \hline 
b$_{4i}'$/a$_5$ & 0 & 0.74\%  \\ \hline 
\multicolumn{3}{|c|}{\thead{Identified dependencies: \\
100\% cases found force-level dependency \\
7.5\% cases failed F-tests, showing no posture dependency \\
20\% cases showed positive posture dependency\\
72.5\% cases showed negative posture dependency}} \\ \hline
\end{tabular}
\end{table} 

\section{Discussion}
\label{ch:Discussion}
We performed a systematic experimental investigation of hip viscoelasticity using NED, a novel rigid robotic interface dedicated to lower limb neuromechanics studies. A position displacement was used as a mechanical perturbation, that enabled us to obtain an accurate estimation of hip stiffness. Viscosity was computed in a second step using a least-square minimization of the linear second order model. The relatively large perturbation amplitude ensured a reliable estimation despite large force measurement noise. We also analysed the influence of the leg, posture, force level and perturbation direction on stiffness and viscosity estimates. The dominant and non-dominant legs exhibited similar values of viscoelasticity, which may not be surprising as the legs are mostly used for the symmetric walking. Sports activities such as playing football might induce some asymmetry, although this could not be studied with the available population. 

Stiffness was found to be slightly larger when estimated from displacement applied in the posterior direction than in the anterior direction. This is probably due to stronger or larger muscles since stiffness is known to vary proportionally to the cross-sectional area of a stretched muscle \cite{Gonzalez1997}, and the quadriceps femoris may be larger than the biceps femoris \cite{Wickiewicz1983}. The study \cite{Koopman2016} estimated hip and knee multi-joint viscoelasticity using an exoskeleton, but could not study the influence of applied force systematically. Using the dedicated NED interface, we could systematically analyse the influence of posture and applied force on the single-joint viscoelastic parameters in a controlled manner. We found that stiffness increases monotonically with the applied limb force, with a relation consistent with previous measurements in the upper limb \cite{Burdet2013}. The stiffness value was found to be slightly influenced by the hip angle, as was previously found in the ankle \cite{Mirbagheri2000}. The viscosity exhibited no clear dependency upon perturbation direction or hip angle, and slightly increases with the applied limb force. The difficulty in identifying viscosity dependencies may originate from its low value relative to stiffness.

The obtained viscoelasticity values we have observed with our subjects population are in the same order as reported in previous studies, although such comparison is limited by the fact that viscoelasticity depends on the individuals. In \cite{Zhang1998}, it was found that knee joint stiffness in the relaxed condition is around 75Nm/rad and viscosity is about 2Nm\,s/rad, and both of these factors increase with muscle contraction. The values we obtained for the hip joint are larger (with stiffness values between 75-318Nm/rad and viscosity 2-21Nm\,s/rad under relaxed condition), as expected as larger muscles are involved. Using the LOPES exoskeleton perturbing the whole leg and indirect position measurement from the serial elastic actuators used in LOPES, \cite{Koopman2016} found stiffness values between 50-220Nm/rad and viscosity between 0.5-10Nm\,s/rad. While being in the same order of magnitude, the difference with the values we have obtained may be in part due to the older population of that study with ages between 67-72 while our young adults were between 21-27.


\bibliographystyle{IEEEtran}
\bibliography{bibliography}

\newpage
\appendix

\section{Mixed effect equation}
\label{ch:app}
The full expansion of Eq.\ref{eq:stiffness_angle} and Eq.\ref{eq:stiffness_angle+force} are, respectively:
\bea
\label{eq:Appendix_stiffness_angle}
\mathbf{K} \!\!\!\!&=&\!\!\! \mathbf{X}\, \boldsymbol{\beta} + \mathbf{Z} \, \boldsymbol{\mu} + \boldsymbol{\varepsilon} \\
\begin{bmatrix} 
K_{11} \\ \vdots \\ K_{1,10} \\ K_{21} \\ \vdots \\ K_{2,10} \\ \vdots \\ K_{51} \\ \vdots \\ K_{5,10} 
\end{bmatrix} \!\!\!\!\!\!&\equiv&\!\!\!\!\!\!
\begin{bmatrix} 
\theta_{11}^2 \!\!\!&\!\!\! \theta_{11} \!\!\!&\!\!\! 1 \\
\vdots \!\!\!&\!\!\! \vdots \!\!\!&\!\!\!\vdots \\ 
\theta_{1,10}^2 \!\!\!&\!\!\! \theta_{1,10} \!\!\!&\!\!\! 1 \\
\theta_{21}^2 \!\!\!&\!\!\! \theta_{21} \!\!\!&\!\!\! 1 \\
\vdots \!\!\!&\!\!\! \vdots \!\!\!&\!\!\!\vdots \\ 
\theta_{2,10}^2 \!\!\!&\!\!\! \theta_{2,10} \!\!\!&\!\!\! 1 \\
\vdots \!\!\!&\!\!\! \vdots \!\!\!&\!\!\!\vdots \\ 
\theta_{51}^2 \!\!\!&\!\!\! \theta_{51} \!\!\!&\!\!\! 1 \\
\vdots \!\!\!&\!\!\! \vdots \!\!\!&\!\!\!\vdots \\ 
\theta_{5,10}^2 \!\!\!&\!\!\! \theta_{5,10} \!\!\!&\!\!\! 1 \\
\end{bmatrix}
\!\!\!\begin{bmatrix}
a_2\\ a_3\\ a_4 \end{bmatrix} \!\! + \!\!
\begin{bmatrix} 
\theta_{11}^2 \, \theta_{11} \, 1 & \!\!\!0\,\,\,\,\,0\,\,\,\,\,0 & \hdots & 0\,\,\,\,\,0\,\,\,\,\,0 \\ 
\vdots & \!\!\!\vdots & & \\ 
\theta_{1,10}^2 \, \theta_{1,10} \, 1 & \!\!\!0\,\,\,\,\,0\,\,\,\,\,0 & & \\ 
0\,\,\,\,\,0\,\,\,\,\,0 & \!\!\!\theta_{21}^2 \, \theta_{21} \, 1 & & \vdots \\ 
 & \!\!\!\vdots & & \\ 
 & \!\!\!\theta_{2,10}^2 \, \theta_{2,10} \, 1 & & \\ 
\vdots & \!\!\!0\,\,\,\,\,0\,\,\,\,\,0 & \ddots & 0\,\,\,\,\,0\,\,\,\,\,0 \\
 & \!\!\!\vdots & & \theta_{51}^2 \, \theta_{51} \, 1 \\
 & \!\!\!\vdots & & \vdots \\
0\,\,\,\,\,0\,\,\,\,\,0 & \!\!\!0\,\,\,\,\,0\,\,\,\,\,0 & & \theta_{5,10}^2 \, \theta_{5,10} \, 1 
\end{bmatrix}\!\!\!
\begin{bmatrix}
b_{21}\\ b_{31} \\ b_{41} \\ b_{22}\\ b_{32}\\ b_{42} \\ \vdots \\ b_{25} \\ b_{35}\\ b_{45}
\end{bmatrix}
\! + \boldsymbol{\varepsilon}
\nonumber
\eea

\bea
\label{eq:Appendix_stiffness_angle+force}
\mathbf{K} \!\!\!\!&=&\!\!\! \mathbf{X}\, \boldsymbol{\beta} + \mathbf{Z}_1 \boldsymbol{\mu}_1 + \mathbf{Z}_2 \boldsymbol{\mu}_2 + \boldsymbol{\varepsilon} \\
\begin{bmatrix} 
\!K_{11}\! \\ \vdots \\ \!K_{16}\! \\ \!K_{21}\! \\ \vdots \\ \!K_{26}\! \\ \vdots \\ \!K_{51}\! \\ \vdots \\ \!K_{56}\!
\end{bmatrix} \!\!\!\!\!\!&\equiv&\!\!\!\!\!\!
\begin{bmatrix} 
\!F_{11} \!\!\!&\!\!\! \theta_{11}^2 \!\!\!&\!\!\! \theta_{11} \!\!\!&\!\!\! 1 \\
\vdots \!\!\!&\!\!\! \vdots \!\!\!&\!\!\! \vdots \!\!\!&\!\!\!\vdots \\ 
\!F_{16} \!\!\!&\!\!\! \theta_{16}^2 \!\!\!&\!\!\! \theta_{16} \!\!\!&\!\!\! 1 \\
\!F_{21} \!\!\!&\!\!\! \theta_{21}^2 \!\!\!&\!\!\! \theta_{21} \!\!\!&\!\!\! 1 \\
\vdots \!\!\!&\!\!\! \vdots \!\!\!&\!\!\! \vdots \!\!\!&\!\!\!\vdots \\ 
\!F_{26} \!\!\!&\!\!\! \theta_{26}^2 \!\!\!&\!\!\! \theta_{26} \!\!\!&\!\!\! 1 \\
\vdots \!\!\!&\!\!\! \vdots \!\!\!&\!\!\! \vdots \!\!\!&\!\!\!\vdots \\ 
\!F_{51} \!\!\!&\!\!\! \theta_{51}^2 \!\!\!&\!\!\! \theta_{51} \!\!\!&\!\!\! 1 \\
\vdots \!\!\!&\!\!\! \vdots \!\!\!&\!\!\! \vdots \!\!\!&\!\!\!\vdots \\ 
\!F_{56} \!\!\!&\!\!\! \theta_{56}^2 \!\!\!&\!\!\! \theta_{56} \!\!\!&\!\!\! 1 \\
\end{bmatrix}
\!\!\!\begin{bmatrix}
a_0'\\ a_2'\\ a_3'\\ a_4' \end{bmatrix} \!\! + \!\!
\begin{bmatrix} 
F_{11} \, 1 & \!0\,\,\,\,\,0 & \hdots & 0\,\,\,\,\,0\!\! \\ 
\vdots & \!\!\!\vdots & & \\ 
F_{16} \, 1 & \!0\,\,\,\,\,0 & & \\ 
0\,\,\,\,\,0\! & \!\!\!F_{21}\,1 & & \vdots \\ 
 & \!\!\!\vdots & & \\ 
 & \!\!\!F_{26}\,1 & & \\ 
\vdots & \!0\,\,\,\,\,0 & \ddots & 0\,\,\,\,\,0\!\! \\
 & \!\!\!\vdots & & F_{51} \, 1 \\
 & \!\!\!\vdots & & \vdots \\
0\,\,\,\,\,0\! & \!0\,\,\,\,\,0 & \hdots & F_{56} \, 1 
\end{bmatrix}\!\!\!
\begin{bmatrix}
\! b_{01}' \! \\ \! b_{11}' \! \\ \! b_{02}' \! \\ \! b_{12}' \! \\ \vdots \\ \! b_{05}' \! \\ \! b_{15}'\! 
\end{bmatrix} 
 \!\!+ \!\!
\begin{bmatrix} 
\theta_{11}^2 \, \theta_{11} \, 1 \!\!&\!\! 0\,\,\,\,\,0\,\,\,\,\,0 \!\!&\!\!  0\,\,\,\,\,0\,\,\,\,\,0 \\ 
\theta_{12}^2 \, \theta_{12} \, 1 \!\!&\!\! 0\,\,\,\,\,0\,\,\,\,\,0 \!\!&\!\!  \vdots \\ 
0\,\,\,\,\,0\,\,\,\,\,0 \!\!&\!\! \theta_{13}^2 \, \theta_{13} \, 1 \!\!&\!\! \vdots  \\ 
\vdots \!\!&\!\! \theta_{14}^2 \, \theta_{14} \, 1 \!\!&\!\! 0\,\,\,\,\,0\,\,\,\,\,0  \\ 
\vdots \!\!&\!\! 0\,\,\,\,\,0\,\,\,\,\,0  \!\!&\!\! \theta_{15}^2 \, \theta_{15} \, 1  \\
0\,\,\,\,\,0\,\,\,\,\,0 \!&\! 0\,\,\,\,\,0\,\,\,\,\,0  \!&\! \theta_{16}^2 \, \theta_{16} \, 1  \\
\theta_{21}^2 \, \theta_{21} \, 1 \!\!&\!\! 0\,\,\,\,\,0\,\,\,\,\,0 \!\!&\!\!  0\,\,\,\,\,0\,\,\,\,\,0 \\ 
\theta_{22}^2 \, \theta_{22} \, 1 \!\!&\!\! 0\,\,\,\,\,0\,\,\,\,\,0 \!\!&\!\!  0\,\,\,\,\,0\,\,\,\,\,0 \\ 
 &\! \vdots  \!&   \\
 0\,\,\,\,\,0\,\,\,\,\,0 \!\!&\!\! 0\,\,\,\,\,0\,\,\,\,\,0  \!\!&\!\! \theta_{55}^2 \, \theta_{55} \, 1  \\
0\,\,\,\,\,0\,\,\,\,\,0 \!\!&\!\! 0\,\,\,\,\,0\,\,\,\,\,0  \!\!&\!\! \theta_{56}^2 \, \theta_{56} \, 1  \\
\end{bmatrix}\!\!\!
\begin{bmatrix}
b_{21}'\\ b_{31}' \\ b_{41}'\\b_{22}'\\ b_{32}' \\ b_{42}'\\b_{23}'\\ b_{33}' \\ b_{43}'
\end{bmatrix}
\!\!+ \boldsymbol{\varepsilon}
\nonumber
\eea

\end{document}